\documentclass[preprint,12pt]{elsarticle}
\usepackage{comment}

\usepackage{amsthm, amsmath, amssymb, bm}
\newtheorem{proposition}{Proposition}

\usepackage{booktabs}

\usepackage{algorithm,algpseudocode}

\begin{document}
\begin{frontmatter}	
\title{Two Parallel PageRank Algorithms via Improving Forward Push}

\author[auth1]{Qi Zhang}
\author[auth1]{Rongxia Tang}
\author[auth1]{Zhengan Yao}
\author[auth2]{Jun Liang \corref{cor1}}
\address[auth1]{Department of Mathematics, Sun Yat-sen University, China}
\address[auth2]{School of Software, South China Normal University, China}
\cortext[cor1]{Corresponding author}

\begin{abstract}
	Initially used to rank web pages, PageRank has now been applied in many fields. With the growing scale of graph, accelerating PageRank computing is urged and designing parallel algorithm is a feasible solution. In this paper, two parallel PageRank algorithms IFP1 and IFP2 are proposed via improving the state-of-the-art Personalized PageRank algorithm, i.e., Forward Push. Theoretical analysis indicates that, IFP1 can take advantage of the DAG structure of the graph, where the dangling vertices improves the convergence rate and the unreferenced vertices decreases the computation amount. As an improvement of IFP1, IFP2 pushes mass to the dangling vertices only once but rather many times, and thus decreases the computation amount further. Experiments on six data sets illustrate that both IFP1 and IFP2 outperform Power method, where IFP2 with 38 parallelism can be at most 50 times as fast as the Power method. 
\end{abstract}

\begin{keyword}
 PageRank, Forward Push, Parallel
\end{keyword}

\end{frontmatter}
\section{Introduction}
S.Brin and L.Page\cite{brin2012reprint,page1999pagerank} proposed PageRank while dealing with the problem of ranking web pages retrieved by Google. PageRank measures the importance of web pages according to the network structure rather than contents. Generally, one web page has higher PageRank value if it is linked by more web pages, or the web pages with a link to it are with higher PageRank value themselves. Nowadays, PageRank's application goes far more beyond Internet\cite{gleich2015pagerank}. We can find it in many fields such as social network analysis, chemistry, molecular biology, sports\cite{brown2017pagerank} and social sciences\cite{Yin2020change}.

In the past decades, plenty of PageRank algorithms were proposed and among which, Power method is the most basic one. Some improvements based on Power method were presented. Kamvar\cite{kamvar2004adaptive} proposed the adaptive method which excluded the vertices had converged from the computation immediately. Haveliwala\cite{haveliwala2003computing} and Kamvar\cite{kamvar2003extrapolation} proposed the extrapolation method which focused on the second largest eigenvalue and made the best use of the previous iteration result. Kamvar\cite{kamvar2003exploiting} proposed the block method by utilizing the network's block structure. Gang\cite{wu2007power} proposed the POWER-ARNOLDI method via introducing Arnoldi-type algorithm into PageRank computation.

Monte Carlo(MC) method \cite{avrachenkov2007monte} is another important PageRank algorithm. MC method simulates random walks on graph and
approximates PageRank value by the probability that random walks terminate at each vertex. Some improvements based on MC method for undirected graph were proposed. Sarma\cite{sarma2013fast} accelerated the walk by stitching short paths into a long one. Luo\cite{2019Distributed,luo2020improved} proposed Radar Push and thus obtained a result with lower variance.

In addition, distributed PageRank algorithms such as Sankaralingam\cite{1210016}, Zhu\cite{zhu2005distributed}, Stergiou\cite{stergiou2020scaling}, Ishii\cite{ishii2010distributed} and \cite{ishii2012web,suzuki2019efficient,charalambous2016totally,suzuki2018distributed,dai2017fully} were proposed. \cite{doi:10.1137/060664331,lin2009computing,zhu2018fast} took advantage of the DAG structure by linear algebra. \cite{duong2012parallel,Lai2017GPGPU} tried computing PageRank on GPU.

Algorithms mentioned above have respective advantages. With the explosive growth in the scale of graph, accelerating PageRank computing is urged. Parallelizing existing algorithms such as Power method\cite{migallon2018parallel} is a feasible solution, however, there stand barriers as below. 
\begin{enumerate}
	\item[(1)] Algorithms based on the Power method converge slowly and can be partially parallel only, since the dependency between different iterations can not be eliminated.
	\item[(2)] Algorithms based on the MC method perform poorly on real application, since large memory space and bandwidth are required.
\end{enumerate}

Forward Push, which is the state-of-the-art Personalized PageRank(PPR) algorithm, has been attracting more attention recently. With initial mass distribution $\bm{p}$, each vertex reserves $1-c$ proportion of the mass receiving from its source vertices, then evenly pushes the remaining $c$ proportion to its target vertices. The mass each vertex obtains in the ending is just the PPR value. Since each vertex executes reserving and pushing operations independently, Forward Push is easy to parallelizing. Designing parallel PageRank algorithm via improving Forward Push is an natural idea. We can obtain the correct PageRank vector on strongly connected graph, however, problems arise when there exist dangling vertices in the graph. Specifically, based on the random walk model, PageRank requires random walk arriving at dangling vertex randomly jump to any vertex, while Forward Push terminates this walk. The slight difference leads that, utilizing Forward Push for PageRank computation forcibly and counting PPR vector as PageRank vector simply, we can either terminate pushing operation at dangling vertices and thus obtain wrong result, or continue pushing mass at dangling vertices to every vertex and thus generate plenty of computation. 

In this paper, we propose parallel PageRank algorithms via improving Forward Push which addressed the problems mentioned above. The contributions are as follows.
\begin{enumerate}
	\item[$(1)$] We reveal that PageRank vector is essentially the probability distribution of mass and can be obtained by Forward Push which terminates pushing mass at dangling vertices. 
	\item[$(2)$] We propose two parallel PageRank algorithms, IFP1 and IFP2, via improving Forward Push. Compared with the Power method, both IFP1 and IFP2 can take advantage of DAG structure of the graph, has higher convergence rate and generates less computation amount. 
	\item[$(3)$] Experimental results on six data sets demonstrate that both IFP1 and IFP2 outperform the Power method, where IFP2 with 38 parallelism can be at most 50 times as fast as. 
\end{enumerate}

The remaining of this paper are as follows. In section 2, we introduce PageRank and Forward Push. In section 3, we firstly give the solution of computing PageRank via Forward Push which terminates pushing mass at dangling vertices, then present IFP1 and corresponding theoretical analysis, and propose IFP2 at last. Some numerical experiments are performed in section 4. We summarize this paper in section 5.

\section{Preliminary}
In this section, PageRank and Forward Push will be briefly introduced. Then the problem arose by dangling vertices when computing PageRank by Forward Push will be detailed. 

\subsection{PageRank}
Given graph $G(V,E)$, where $V = \{v_{1}, v_{2}, \cdots, v_{n}\}$, $E=\{(v_{i},v_{j}): i,j=1,2,\cdots,n\}$ and $\vert E \vert = m$. Denote by $\bm{A}=(a_{ij})_{n \times n}$ the adjacency matrix, where $a_{ij}$ is 1 if $(v_{j},v_{i}) \in E$ and 0 else. Denote by $\bm{P}=(p_{ij})_{n \times n}$ the probability transition matrix, where
\begin{center}
	$
	p_{ij} = \left\{ 
	\begin{array}{ll}
		a_{ij} / \sum \limits_{i=1}^{n} {a_{ij}}, & \textrm{if $\sum \limits_{i=1}^{n} {a_{ij}} \neq 0$},\\
		0, & \textrm{else}.
	\end{array}
	\right.
	$
\end{center}
We call vertex without in-link the unreferenced vertex, vertex without out-link the dangling vertex. Delete unreferenced (dangling) vertex, the newly generating unreferenced (dangling) vertex is called the weak unreferenced (dangling) vertex. Let $\bm{d}=(d_{1},d_{2},...,d_{n})^{T}$, where $d_{i}$ is 1 if $v_{i}$ is dangling vertex and 0 else. Let $\bm{p}=(p_{1},p_{2},\cdots,p_{n})^{T}$ denote the $n$-dimensional probability distribution. Let $\bm{P}^{'}=\bm{P}+\bm{p}\bm{d}^{T}$ and $\bm{P}^{''}=c\bm{P}^{'}+(1-c)\bm{p}\bm{e}^{T}$, where $c\in(0,1)$ is damping factor and $\bm{e}=(1,1,...,1)^{T}$. Denote by $\bm{\pi}$ the PageRank vector. According to\cite{Berkhin2005A}, let $\bm{p}=\frac{\bm{e}}{n}$, one of PageRank's definitions is 
\begin{equation}\label{equa1}
  \bm{\pi}=\bm{P}^{''}\bm{\pi},\pi_{i}>0,\sum \limits_{i=1}^{n}\pi_{i}=1.
\end{equation}

PageRank can be interpreted from many perspectives. Based on random walk model, a random walk starts at random vertex, with probability $c$ walks according to the graph, and with probability $1-c$ terminates, then $\bm{\pi}$ is the probability distribution of random walk terminating at each vertex. When $\bm{p} \neq \frac{\bm{e}}{n}$, PageRank converts to Personalized PageRank(PPR), and $\bm{p}$ is called the personalized vector. PageRank is a special case of PPR. 

It should be noted that, the graph which rules the random walk is actually the graph corresponding to $\bm{P}^{'}$ but rather the original one. The difference between these two is mainly on the dangling vertices, specifically, the former artificially links each dangling vertex to every vertices while the latter does nothing. As a consequence, when a random walk arriving at the dangling vertices, the former requires with probability $c$ randomly choosing a target vertex and continuing the random walk, while the latter means terminating. Graph corresponding to $\bm{P}^{'}$ contains more extra edges. From the perspective of computing, the existing of dangling vertices always generates more computation when walking according to $\bm{P}^{'}$, and that is why directly utilizing Forward Push in computing PageRank is costly. 

\subsection{Forward Push}
Forward Push is the state-of-the-art PPR algorithm. As described in Algorithm \ref{forwardpush}, with initial mass distribution $\bm{p}$, each vertex does the following two: (1)reserve $1-c$ proportion of mass it has received; (2)evenly push the remaining $c$ proportion to its target vertices. While Forward Push running, the pushing mass, i.e., the mass needing to push to the target vertices decreases and the reserved mass increases. The algorithm finishes when there's no vertex holds more than the pre-defined $\xi$ pushing mass. The reserved mass of each vertex is just its PPR value. 
\begin{algorithm}[htbp]
	\scriptsize
	\begin{algorithmic}[1]
		\Require{
			\\$G(V,E)$:The graph;
			\\$c$:The damping factor;
			\\$\bm{p}$:The personalized vector;
			\\$\xi$:The tolerance of error.}
		\Ensure{\\$\bm{\pi}$:Personalized PageRank vector.}
		\State
		\State{Each vertex $v_{i}$ maintains a data structure $\langle \overline{\pi}_{i},h_{i} \rangle$.}
		\State{Initially set $\overline{\pi}_{i}=0$, $h_{i}=p_{i}$.}
		\State
		\While{There exists vertex $v_{i}$ satisfying $h_{i} > \xi$}
		\State{$\overline{\pi}_{i} += (1-c)h_{i}$;}
		\For{$v_{j} \in D(v_{i})$}\Comment{[$D(v_{i})$ are target vertices of $v_{i}$.]}
		\State{$h_{j} += \frac{c h_{i}}{deg(v_{i})}$;}
		\EndFor
		\State{$h_{i}=0$;}
		\EndWhile
		\State{$\pi_{i}=\overline{\pi}_{i}$.}
	\end{algorithmic}
	\caption{\textbf{Forward Push}}
	\label{forwardpush}
\end{algorithm}

Forward Push is convenient to parallelizing since each vertex executes reserving and pushing operation independently. Previous works seldom detailed how to address mass on the dangling vertices. As a special case of PPR, PageRank can be obtained by Forward Push when the graph is strongly connected. However, for graph containing dangling vertices, Forward Push can either
\begin{enumerate}
	\item [(1)] terminate pushing the mass on dangling vertices, and thus obtain an incorrect result; 
	\item [(2)] continue pushing mass on dangling vertices evenly to every vertex, and thus generate plenty of computation. 
\end{enumerate}
Moreover, in real computing environment, if the scale of graph is large enough, the mass each vertex getting from the dangling vertices will be so small that the result may loss precision. Graphs abstracted from reality always contain large proportion of dangling vertices. To design parallel PageRank algorithm based on Forward Push, the problem arose by dangling vertices needs to be addressed firstly.

\section{Algorithm and convergence analysis}
In this section, we firstly demonstrate that PageRank vector is essentially a distribution of mass, and it is feasible to compute PageRank via Forward Push that needs not push mass on dangling vertices. Then we propose the parallel PageRank algorithm IFP1 by improving Forward Push and present theoretical analysis as well. At last, we propose IFP2, the improvement of IFP1.

\subsection{Computing PageRank vector via Forward Push}
By expanding Formula (\ref{equa1}), we have
\begin{equation}\label{equa2}
	(\bm{I}-c\bm{P}^{'})\bm{\pi}=(1-c)\bm{p}.
\end{equation}
Since $\rho(c\bm{P}^{'})<1$, $(\bm{I}-c\bm{P}^{'})^{-1}=\sum\limits_{r=0}^{\infty}(c\bm{P}^{'})^{r}$, it follows that 
\begin{equation}\label{equa3}
	\bm{\pi}=(1-c)\sum \limits_{r=0}^{\infty}(c\bm{P}^{'})^{r}\bm{p}.
\end{equation}
Formula (\ref{equa3}) is just the algebraic form of Forward Push, the masses on dangling vertices are pushed to every vertex according to $\bm{p}$.

By expanding \eqref{equa2}, we have 
\begin{center}
	$(\bm{I}-c\bm{P})\bm{\pi}=(c\bm{d^{T}\pi}+1-c)\bm{p}$.
\end{center}
Since $\bm{I}-c\bm{P}$ is invertible , it follows that 
\begin{center}
	$\bm{\pi}=(c\bm{d^{T}\pi}+1-c)(\bm{I}-c\bm{P})^{-1}\bm{p}$.
\end{center}
Let $\gamma=c\bm{d^{T}\pi}+1-c$, it follows that 
\begin{center}
	$\bm{\pi}=\gamma(\bm{I}-c\bm{P})^{-1}\bm{p}$.
\end{center}
Since $1=\bm{e^{T}\pi}=\gamma \bm{e^{T}}(\bm{I}-c\bm{P})^{-1}\bm{p}$, we have  
\begin{center}
	$\gamma=\frac{1}{\bm{e^{T}}(\bm{I}-c\bm{P})^{-1}\bm{p}}$.
\end{center}
Since $\pi_{i}=\gamma \bm{e_{i}^{T}}(\bm{I}-c\bm{P})^{-1}\bm{p}$, where $\bm{e}_{i}$ is $n$-dimensional vector with the $i_{th}$ element is 1 and 0 others, it follows that 
\begin{center}
	$\pi_{i}=\frac{\bm{e_{i}^{T}}(\bm{I}-c\bm{P})^{-1}\bm{p}}{\bm{e^{T}}(\bm{I}-c\bm{P})^{-1}\bm{p}}$.
\end{center}
Since $(\bm{I}-c\bm{P})^{-1}=\sum \limits_{r=0}^{\infty}(c\bm{P})^{r}$, we have 
\begin{equation}\label{equa4}
	\pi_{i}=\frac{\bm{e_{i}^{T}}\sum \limits_{r=0}^{\infty}(c\bm{P})^{r}\bm{p}}{\bm{e^{T}}\sum \limits_{r=0}^{\infty}(c\bm{P})^{r}\bm{p}}.
\end{equation}
$\sum \limits_{r=0}^{\infty}(c\bm{P})^{r}\bm{p}$ is the algebraic form of Forward Push as well, it is different from Formula (\ref{equa3}) on two aspects: (1)it terminates pushing mass on the dangling vertices; (2) it reserves 100\% but not $1-c$ proportion of the mass.

Formula (\ref{equa4}) demonstrates that PageRank vector is essentially a distribution of reserving mass. The proportion each vertex reserves has no effect on the final result, however, if each vertex reserves 100\% proportion of mass, there's no need for dangling vertices executing reserving operation. Moreover, it is $\bm{P}$ but rather $\bm{P}^{'}$ that ruled the process of Forward Push, the mass on dangling vertices need not to be pushed any more. Formula (\ref{equa4}) implies a solution of addressing the problem arose by dangling vertices. 

\subsection{IFP1}
The remaining issue is parallelizing. Restricted by the computing resource, assigning exclusive thread for every vertex is infeasible. We can generate some threads and assign vertices to them. Then IFP1 is proposed as Algorithm \ref{IFP1}. 
\begin{algorithm}[t]
	\scriptsize
	\begin{algorithmic}[1]
		\Require{\\$K$:The number of threads;
			\\$\xi$:The lower bound of mass.}
		\Ensure{\\$\bm{\pi}$:PageRank vector.}
		\State{Each vertex $v_{i}$ maintains a data structure $\langle \overline{\pi}_{i},h_{i} \rangle$.}
		\State{Assign non-dangling vertices to $K$ threads, denote by $S_{j}$ the set of vertices belonging to thread $j$.}
		\State{Initially set $\overline{\pi}_{i}=0$, $h_{i}=1$.}
		\State{Invoke $K$ Calculations and Management;}\Comment{[The $K$ Calculations and Management do in parallel.]}
		\State{Calculate $\pi$ following $\pi_{i}=\frac{\overline{\pi}_{i}+h_{i}}{\sum\limits_{i=1}^{n}(\overline{\pi}_{i}+h_{i})}$ while the Management terminates.}
		\State
		\Function{Calculation}{$j$}
		\While {1}
		\For{$v_{i} \in S_{j}$}
		\If{$h_{i}>\xi$}
		\State{$\overline{\pi}_{i}=\overline{\pi}_{i}+h_{i}$;}
		\For{$u \in D(v_{i})$}\Comment{[$D(v_{i})$ is the set of target vertices of $v_{i}$.]}
		\State{$h_{u}=h_{u}+\frac{ch_{i}}{deg(v_{i})}$;}
		\EndFor
		\State{$h_{i}=0;$}
		\EndIf
		\EndFor
		\EndWhile
		\EndFunction
		\State
		\Function{Management}{}
		\While{There exists non-dangling vertex satisfying $h_{i}>\xi$}
		\EndWhile
		\State{Terminate all the $K$ Calculations.}
		\EndFunction
	\end{algorithmic}
	\caption{IFP1}
	\label{IFP1}
\end{algorithm}

While IFP1 running, thread $j$ circularly scans $S_{j}$, any vertex $v_{i}$ satisfying $h_{i}>\xi$ will be processed. The mass received from its source vertices are reserved by 100\% proportion, and then pushed to its target vertices by $c$ proportion. On the whole, the reserved mass increases and pushing mass decreases. The reserved mass of unreferenced vertices and weak unreferenced vertices stay steadily after several iterations, i.e., these vertices get converged. If there exists no non-dangling vertex holds more than the predefined $\xi$ pushing mass, IFP1 gets converged. IFP1 is similar to Forward Push except the following:
\begin{enumerate}
	\item [(1)] IFP1 is parallel while Forward Push is serial;   
	\item [(2)] IFP1 processes non-dangling vertices only while Forward Push addresses all of them;   
	\item [(3)] IFP1 reverses 100\% proportion of the mass while Forward Push reserves $1-c$ proportion. 
\end{enumerate}
It should be noted that, in multi-thread environment, the addition on $h_{i}$ must be atomic and thus data race may occur.

\subsection{Algorithm analysis}
We analyse IFP1 from three aspects, the convergence rate, error and computation amount.

\subsubsection{Convergence rate}
Convergence rate relates to iteration rounds, however, as a parallel algorithm IFP1 in fact has no iteration. We define one iteration of IFP1 as that the threads finish scanning through the whole vertices. Since whether processes the dangling vertices, the proportion of reserving mass and the parallelism have no effect on the convergence rate, for convenience, we assume that (1)IFP1 addresses all of the vertices; (2)the proportion of reserving mass is still $1-c$; (3)the parallelism is 1. 

Denote by $\overline{\bm{\pi}}^{I}(t)$ the mass having been reserved at the beginning of the $t_{th}$ iteration, by $\overline{\bm{\pi}}^{R}(t)$ the mass needing to push at the beginning of the $t_{th}$ iteration, then it follows that 
\begin{center}
	$\vert \vert \overline{\bm{\pi}}^{I}(t)+\overline{\bm{\pi}}^{R}(t) \vert\vert_{1} \leq n$, 
\end{center}
where $\overline{\bm{\pi}}^{I}(0)=\bm{0}$, 
$\overline{\bm{\pi}}^{R}(0)=\bm{e}$, 
$\lim \limits_{t \to \infty}\frac{\overline{\bm{\pi}}^{I}(t)}{\vert\vert \overline{\bm{\pi}}^{I}(t) \vert\vert_{1}}=\bm{\pi}$,
$\lim \limits_{t \to \infty}\overline{\bm{\pi}}^{R}(t)=\bm{0}$, 
$\vert\vert \overline{\bm{\pi}}^{I}(t)\vert\vert_{1}$ increases monotonously on $t$, and  $\vert\vert\overline{\bm{\pi}}^{R}(t)\vert\vert_{1}$ decreases monotonously on $t$. The convergence rate can be measured by $\frac{\vert\vert \overline{\bm{\pi}}(t) \vert\vert_{1}}{\vert\vert \overline{\bm{\pi}}(t+1) \vert\vert_{1}}$. 

At the beginning of the $t_{th}$ iteration, some vertices having been converged and denote them by $V_{U}(t)$. Denote by $V_{D}$ the set of dangling vertices, it follows that 
\begin{center}
	$\vert\vert \overline{\bm{\pi}}^{R}(t) \vert\vert_{1}=\sum \limits_{v_{i} \in V_{U}(t)}\overline{\pi}^{R}_{i}(t) + \sum \limits_{v_{i} \in V_{1}(t)}\overline{\pi}^{R}_{i}(t) + \sum \limits_{v_{i} \in V_{2}(t)}\overline{\pi}^{R}_{i}(t)$,  
\end{center}
where $V_{1}(t)=V_{D}-V_{U}(t)$ and $V_{2}(t)=V-V_{U}(t)-V_{D}$. 

During the $t_{th}$ iteration, the reserved mass of $V_{U}(t)$ stay constant; the mass needing to push of $V_{1}(t)$ decrease to 0; the sum of mass needing to push of $V_{2}(t)$ decrease by $1-c$ proportion, thus it follows that 
\begin{center}
	$\vert\vert \overline{\bm{\pi}}^{R}(t+1) \vert\vert_{1}=\sum \limits_{v_{i} \in V_{U}(t)}\overline{\pi}^{R}_{i}(t) + c\sum \limits_{v_{i} \in V_{2}(t)}\overline{\pi}^{R}_{i}(t)$.
\end{center}
Vertices belonging to $V_{U}(t)$ have been converged, we have 
\begin{center}
	$\sum \limits_{v_{i} \in V_{U}(t)}\overline{\pi}^{R}_{i}(t) < \vert  V_{U}(t) \vert \xi$. 
\end{center}
Assuming that $\xi$ is sufficiently small that $\vert  V_{U}(t) \vert \xi \to 0$, then 
\begin{center}
$\frac{\vert\vert\overline{\bm{\pi}}^{R}(t+1)\vert\vert_{1}}{\vert\vert\overline{\bm{\pi}}^{R}(t)\vert\vert_{1}}=
\frac{c\sum_{v_{i} \in V_{2}(t)}\overline{\pi}^{R}_{i}(t)}{\sum_{v_{i} \in V_{1}(t)}\overline{\pi}^{R}_{i}(t)+\sum_{v_{i} \in V_{2}(t)}\overline{\pi}^{R}_{i}(t)}$.  
\end{center}
Let $\alpha(t)=\frac{\sum_{v_{i} \in V_{2}(t)}\overline{\pi}^{R}_{i}(t)}{\sum_{v_{i} \in V_{1}(t)}\overline{\pi}^{R}_{i}(t)+\sum_{v_{i} \in V_{2}(t)}\overline{\pi}^{R}_{i}(t)}$, then
\begin{center}
$\frac{\vert\vert\overline{\bm{\pi}}^{R}(t+1)\vert\vert_{1}}{\vert\vert\overline{\bm{\pi}}^{R}(t)\vert\vert_{1}}=c\alpha(t)$.  
\end{center}
Since $V_{1}(t) \cup V_{2} = V-V_{U}(t)$, we have 
\begin{center}
$\alpha(t)=\frac{\sum_{v_{i} \in V-V_{U}(t)- V_{D}}\overline{\pi}^{R}_{i}(t)}{\sum_{v_{i} \in V-V_{U}(t)}\overline{\pi}^{R}_{i}(t)}$. 
\end{center}
$\alpha(t)$ is the proportion of pushing mass on non-dangling vertices among the total pushing mass at the beginning of the $t_{th}$ iteration. 

Generally, $\alpha(t)$ negatively correlates with the proportion of dangling vertices. While IFP1 running, the amount of non-converged vertices decreases, then the proportion of dangling vertices among non-converged vertices increases, and thus $\alpha(t)$ decreases. Let $\alpha=\max \limits_{t \geq 1} \left\{\alpha(t)\right\}$ and $\lambda=\alpha c$, then 
\begin{center}
$\min \limits_{t \geq 1} \left\{\frac{\vert\vert\overline{\bm{\pi}}^{R}(t)\vert\vert_{1}}{\vert\vert\overline{\bm{\pi}}^{R}(t+1)\vert\vert_{1}}\right\}=
\frac{1}{\max \limits_{t \geq 1}\left\{\alpha(t)\right\} c}= 
\frac{1}{\alpha c}=
\lambda^{-1}$.  
\end{center}
$\lambda^{-1}$ can be viewed as IFP1's convergence rate. Generally, the larger proportion of dangling vertices a graph has, the higher convergence rate IFP1 is. The iterations IFP1 takes to get converged under the predefined threshold $\xi$ can be estimated by 
\begin{center}
$\vert\vert\overline{\bm{\pi}}^{R}(T)\vert\vert_{1}=n\prod \limits_{t=0}^{T-1}c\alpha(t)=\sum \limits_{v_{i} \in V_{1}(t)}{\overline{\pi}^{R}_{i}(T)} + \sum \limits_{v_{i} \in V_{2}(t)}{\overline{\pi}^{R}_{i}(T)}$.
\end{center}
Since IFP1 gets converged when $\vert \vert \bm{\pi}_{R}(t)\vert \vert_{\infty} < \xi$ and mass on $V_{1}(t)$ decrease to 0 during the $t_{th}$ iteration, one of the sufficient conditions is
\begin{center}
	$n\prod \limits_{t=0}^{T-1}c\alpha(t) \leq n(\lambda)^{T} < \vert V-V_{U}(t)-V_{D}\vert \xi$,   
\end{center}
and thus 
\begin{center}
	$T>\log_{\lambda}{\xi}+\log_{\lambda}{\frac{\vert V-V_{U}(T)-V_{D} \vert}{n}}$. 
\end{center}
Generally, $\vert V-V_{U}(T)-V_{D} \vert=O(n)$, we have 
\begin{equation}\label{equa7}
T=O(\log_{\lambda}{\xi}).  
\end{equation}

\subsubsection{Error}
It is infeasible to estimate the error of each vertex since the complication of graph structure. We discuss the effect of pre-defined threshold $\xi$ on relative error $ERR(\xi)$ from the whole. Assuming $\vert\vert \overline{\bm{\pi}}^{R}(t)\vert\vert_{1}=n\lambda^{t}$, then 
\begin{center}
$\vert\vert \overline{\bm{\pi}}^{I}(T+t)-\overline{\bm{\pi}}^{I}(T)\vert\vert_{1} \leq n\vert\lambda^{T}-\lambda^{T+t}\vert=(1-\lambda^{t})n\lambda^{T}$. 
\end{center}
Let $t \rightarrow \infty$, then we can obtain that
\begin{equation}\label{equa8}
	ERR(\xi)=\lim\limits_{t \rightarrow \infty}\frac{\vert\vert \bm{\pi}^{I}(T+t)-\bm{\pi}^{I}(T)\vert\vert_{1}}{\vert\vert\bm{\pi}^{I}(T+t)\vert\vert_{1}}<\lim\limits_{t \rightarrow \infty}\frac{\vert\lambda^{T}-\lambda^{T+t}\vert}{1-\lambda^{T+t}}=\xi. 
\end{equation}
Formula (\ref{equa8}) can be viewed as an estimation of relative error under the predefined threshold $\xi$. It is to be noted that the error analysis mentioned above is rough, it considers no any factor of graph structure. 

\subsubsection{Computation amount}
The computation of IFP1 mainly consists by additions and productions. Denote by $m(t)$ the computation amount of the $t_{th}$ iteration, and by $M$ the total computation amount for IFP1 getting converged. Since only vertices belonging to $V-V_{U}(t)$ generate arithmetic operations during the $t_{th}$ iteration, it follows that 
\begin{center}
	$m(t)=\sum_{v_{i} \in V-V_{U}(t)} (deg(v_{i})+1)$.
\end{center}
Let $\beta(t)=\frac{m(t+1)}{m(t)}$. $\beta(t)$ indicates the decreasing rate of computation amount. Unreferenced vertices and weak unreferenced vertices gradually exit computing while IFP1 running, thus $\beta(t) \leq 1$ and 
\begin{center}
	$m+n=m(0) \geq m(1) \geq m(2) \geq \cdots \geq m(T)$. 
\end{center}
Let $\beta=\max \limits_{1 \leq t \leq T} \left\{\beta(t)\right\}$. Assuming that $\beta < 1$, i.e., there exist vertices getting converged at each iteration, we have  
\begin{center}
	$M=\sum \limits_{t=0}^{T} m(t)=\sum \limits_{t=0}^{T} \sum_{v_{i} \in V-V_{U}(t)}(deg(v_{i})+1) < (m+n)\frac{1-\beta^{T}}{1-\beta}$, 
\end{center}
where $T=O(\log_{\lambda}{\xi})$. 

As a summary, it is clear that IFP1 could take advantage of DAG structure, where the dangling vertices improve the convergence rate and the unreferenced vertices lower the computation amount. Compared with Power method, on graph containing DAG structure, IFP1 requires less iterations to get converged and generates less computations at each iteration, thus is faster.

\subsection{IFP2}
IFP1 requires no the dangling vertices executing reserving operation, however, that pushing mass to dangling vertex from its source vertices is still needed and sometimes will be executed many times. Specifically, denote by $v_{d}$ a dangling vertex and by $S(v_{d})$ the set of $v_{d}$'s source vertices, each time $v_{i}$ satisfies $h_{i}>\xi$, the pushing operation from $v_{i}$ to $v_{d}$ will be executed. On the other hand, we have 
\begin{center}
$\sum \limits_{t=0}^{T} \overline{\pi}_{d}^{R}(t)= \sum \limits_{v_{i} \in S(v_{d})} \frac{c}{deg(v_{i})} \sum \limits_{t=0}^{T}\overline{\pi}_{i}^{R}(t)$, 
\end{center}
where $\sum \limits_{t=0}^{T}\overline{\pi}_{i}^{R}(t)$ is the reserved mass of $v_{i}$. That implies the mass on dangling vertices is completely determined by their source vertices. If we do not push mass to the the dangling vertices from their source vertices initially, but execute these operations after all the non-dangling vertices getting converged, then only once pushing operation is sufficient. Motivated by this, IFP2, an improvement of IFP1, is proposed as Algorithm \ref{IFP2}. 

\begin{algorithm}[htbp]
\scriptsize
\begin{algorithmic}[1]
	\Require{\\$K$:The number of threads;
			\\$\xi$:The lower bound of mass.}
	\Ensure{\\$\bm{\pi}$:PageRank vector.}
	\State
	\State{Preprocess the graph data and invalid edges between dangling vertices and their source vertices.}
	\State{Each vertex $v_{i}$ maintains a data structure $\langle \overline{\pi}_{i},h_{i} \rangle$.} 
	\State{Assign vertices to $K$ threads, denote by $S^{1}_{j}$ and $S^{2}_{j}$ the set of non-dangling vertices and dangling vertices belonging to thread $j$ respectively.}
	\State{Initially set $\overline{\pi}_{i}=0$, $h_{i}=1$.}
	\State{Invoke $K$ Calculations and Management;}\Comment{[The $K$ Calculations and Management do in parallel.]}
	\State{Calculate $\pi$ following $\pi_{i}=\frac{\overline{\pi}_{i}}{\sum\limits_{i=1}^{n}\overline{\pi}_{i}}$ while the Management terminates.}
	\State
	\Function{Calculation}{$j$}
	\While{1}
	\If{$CTRL$} \Comment{[$CTRL$ is bool and initially true.]}
	\State{Phase1:}
	\For{$v_{i} \in S^{1}_{j}$}
	\If{$h_{i}>\xi$}
	\State{$\overline{\pi}_{i}=\overline{\pi}_{i}+h_{i}$;}
	\For{$u \in D(v_{i})$}\Comment{[$D(v_{i})$ is the set of target vertices of $v_{i}$.]}
	\State{$h_{u}=h_{u}+\frac{ch_{i}}{deg(v_{i})}$;}
	\EndFor
	\State{$h_{i}=0;$}
	\EndIf
	\EndFor
    \Else
    \State{Phase2:}
	\If{$STS$} \Comment{[$STS$ is bool and initially true.]}
	\For{$v_{i} \in S^{2}_{j}$}
	\State{$\overline{\pi}_{i}=h_{i}$;}
	\For{$u \in S(v_{i})$} \Comment{[$S(v_{i})$ is the set of source vertices of $v_{i}$.]}
	\State{$\overline{\pi}_{i}=\overline{\pi}_{i}+\frac{c\overline{\pi}_{u}}{deg(u)}$;}
	\EndFor
	\State{$h_{i}=0$;}
	\EndFor
	\State{$STS=false$;}
	\EndIf
	\EndIf
	\EndWhile
	\EndFunction
	\State
	\Function{Management}{}
	\While{There exists non-dangling vertex satisfying $h_{i}>\xi$}
	\EndWhile
	\State{$CTRL=false$;}
	\While{There exists dangling vertex satisfying $h_{i}>\xi$}
	\EndWhile
	\State{Terminate all the $K$ Calculations.}
	\EndFunction
\end{algorithmic}
\caption{IFP2}
\label{IFP2}
\end{algorithm}

Different from IFP1, IFP2 needs to address all of the vertices, both non-dangling vertices and dangling vertices are assigned to the $K$ threads. IFP2 contains two phases, where phase1 addresses the non-dangling vertices and phase2 addresses the dangling vertices. Phase1 is similar to IFP1 except that no mass is pushed to the dangling vertices. While phase1 finished, all of the non-dangling vertices get converged, phase2 starts and pushes mass to the dangling vertices. Compared with IFP1, phase2 executes only once and thus some pushing operation corresponding to dangling vertices are saved. Specifically, during the $t_{th}$ iteration, there are $\sum \limits_{v \in V_{D}-V_{U}(t)} \vert S(v)\vert$ mass pushing operations saved, thus the total saved operations is $\sum \limits_{t=0}^{T}\sum \limits_{v \in V_{D}-V_{U}(t)} \vert S(v)\vert$. Generally, the higher proportion of edges corresponding to dangling vertices the graph contains, the less computation IFP2 generates. 

It should be noted that, IFP2 needs some preprocessing, before phase1 running, edges corresponding to dangling vertices should be invalidated. These work may increase the time consumption, however, based on appropriate data structure, the preprocessing cost is at most $o(\sum \limits_{v \in V_{D}} \vert S(v)\vert)$. Experiments in the following will show that, the invaliding operations almost add no extra CPU time consumption.

\section{Experiment}\label{sec5}
In this section, both IFP1 and IFP2 are demonstrated experimentally. Firstly, we introduce the computing environment and indexes to be used. Then, the convergence of IFP1 and IFP2 are illustrated, and the comparison with Power method as well. At last, the results of IFP2 with different threads is elaborated.

\subsection{Experiment Setting}
All algorithms of this experiment are implemented with C++ on serves with Intel(R) Xeon(R) Silver 4210R CPU 2.40GHz 40 processors and 190GB memory. The operation system is Ubuntu 18.04.5 LTS. Six data sets are illustrated in table \ref{table1}, where $n$, $m$, $n_{d}$, $m_{d}$ and $deg=\frac{m}{n}$ represent the number of vertices, the number of edges, the number of dangling vertices, the number of edges corresponding to the dangling vertices and the average degree respectively. The CPU time consumption $T$ and max relative error $ERR=\max \limits_{v_{i} \in V} \frac{\vert \overline{\pi}_{i}- \pi_{i} \vert}{\pi_{i}}$ are taken to estimate the algorithms, where  The true PageRank value $\pi_{i}$ is obtained by Power method at the $210_{th}$ iteration. The damping factor $c=0.85$. 

\begin{table}[htbp]
\footnotesize
\centering
\begin{tabular}{l l l l l l}
	\toprule
	Data Sets            &$n$         &$m$         &$n_{d}$        &$m_{d}$      &$deg$    \\
	\midrule
	web-Stanford         &281903      &2312497     &172         &410       &8.21           \\
	Stanford-Berkeley    &683446      &7583376     &68062       &994368    &11.1           \\
	web-Google           &916428      &5105039     &176974      &325725    &5.57           \\
	in-2004              &1382908     &16917053    &282306      &1006484   &12.23          \\
	soc-LiveJournal1     &4847571     &68993773    &539119      &2092984   &14.23          \\
	uk-2002              &18520343    &298113762   &2760973     &16029357  &16.10          \\
	\bottomrule
\end{tabular}
\caption{Data Sets}
\label{table1}
\end{table}

\subsection{Convergence}
\begin{figure}[htbp]
	\centering
	\includegraphics[width=0.32\textwidth,height=0.20\textheight]{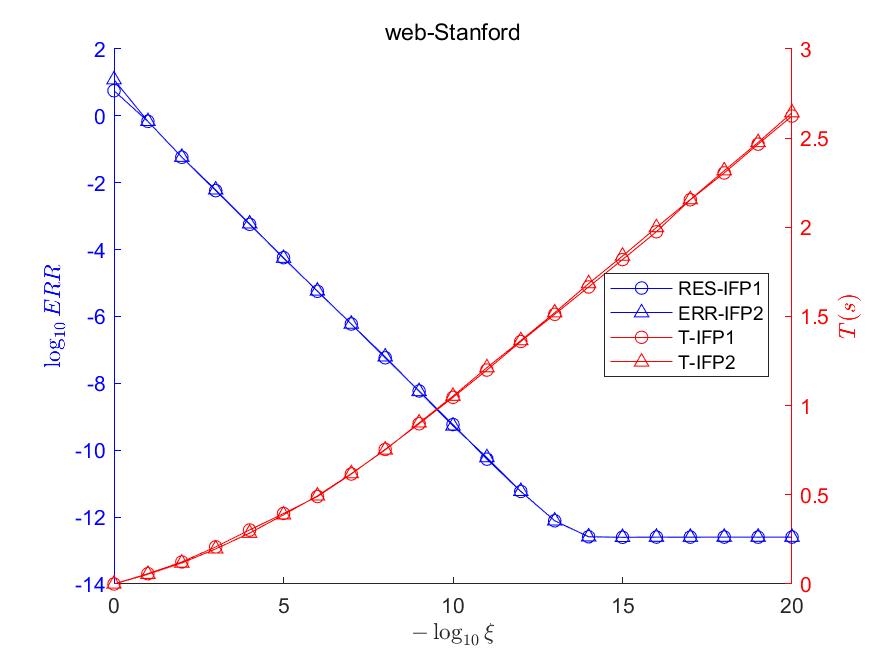}
	\includegraphics[width=0.32\textwidth,height=0.20\textheight]{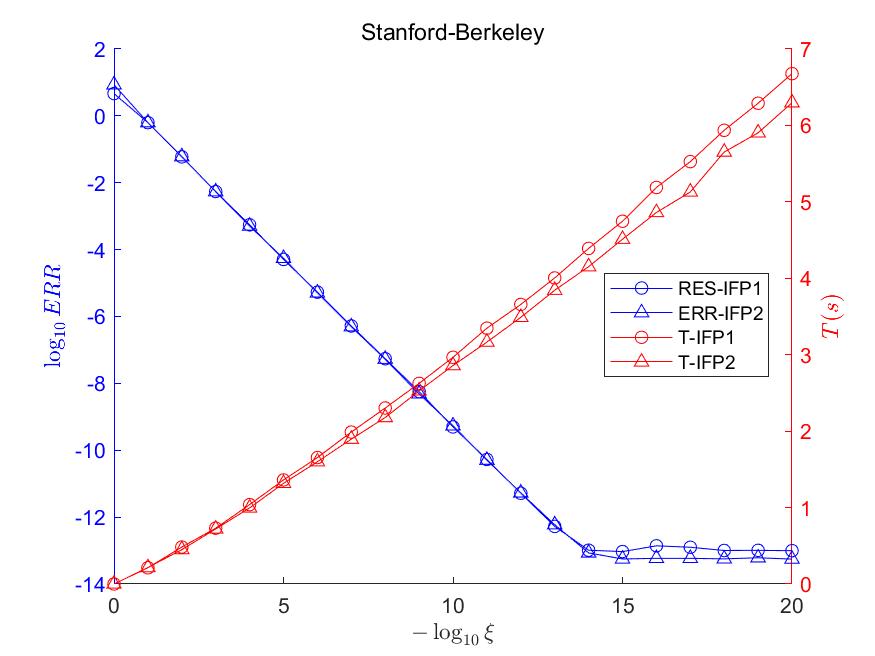}
	\includegraphics[width=0.32\textwidth,height=0.20\textheight]{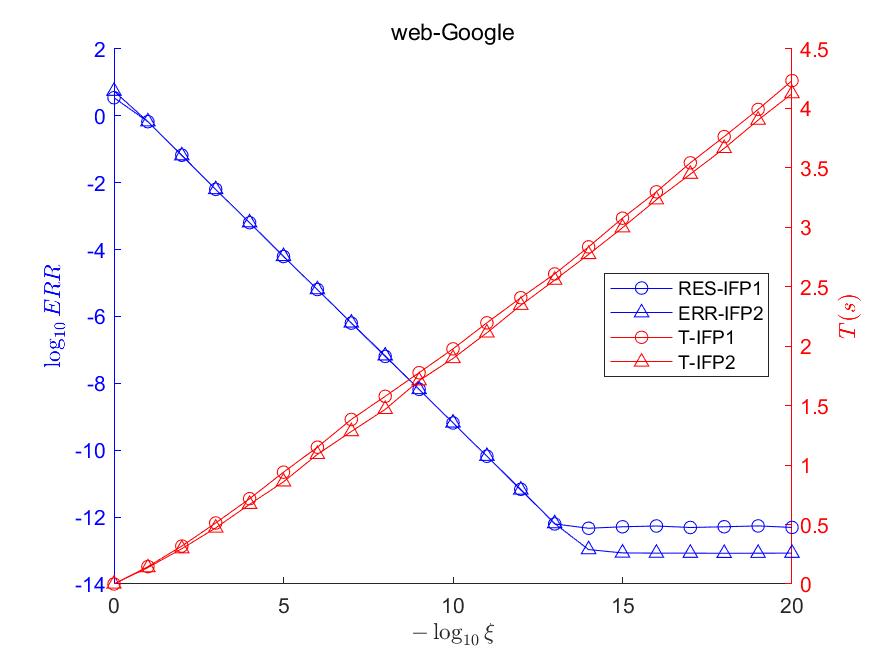}
	\includegraphics[width=0.32\textwidth,height=0.20\textheight]{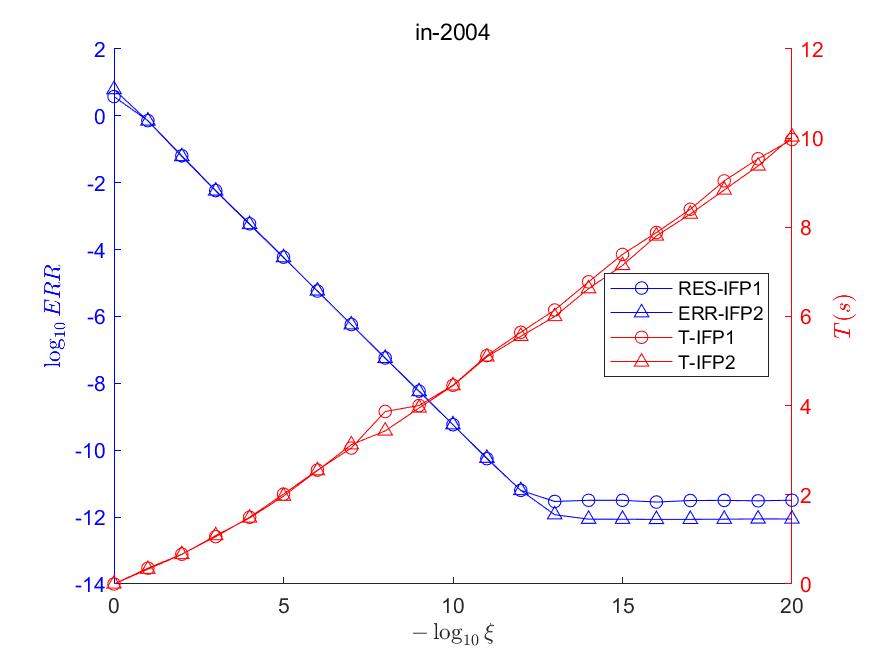}
	\includegraphics[width=0.32\textwidth,height=0.20\textheight]{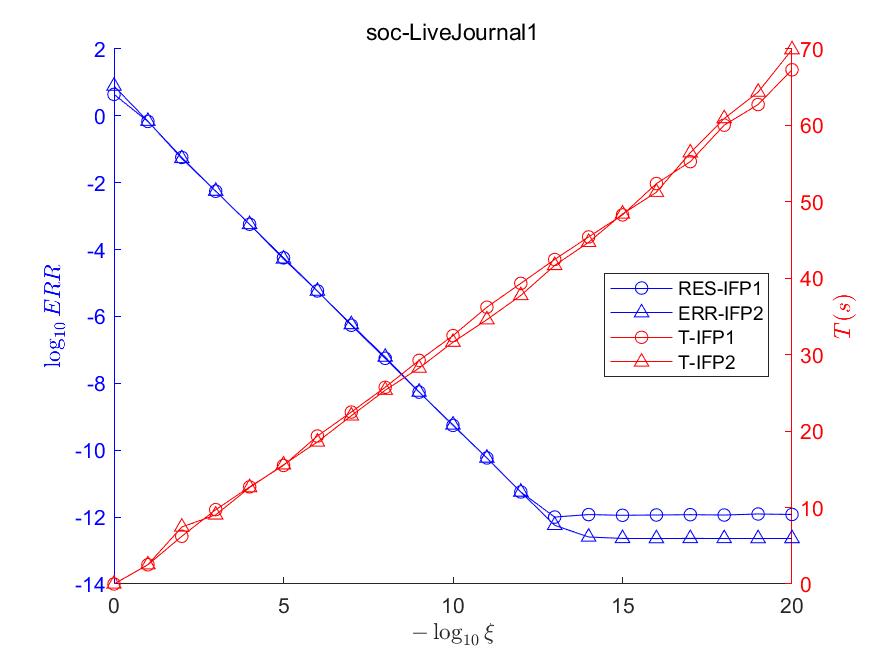}
	\includegraphics[width=0.32\textwidth,height=0.20\textheight]{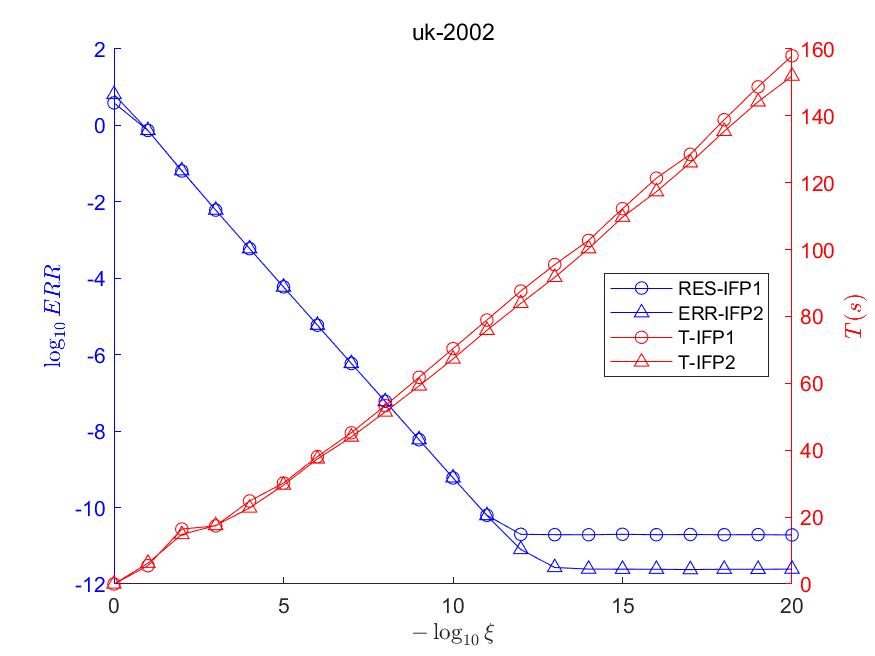}
	\caption{$\xi$ Versus $RES$ and $T$}
	\label{fig1}
\end{figure}

The relation of $ERR$, $T$ and $\xi$ on six data sets is illustrated in Figure \ref{fig1}. 
\begin{enumerate}
	\item[(1)] The blue lines show that $ERR$ has a positive linear relation with $\xi$, which is consistent with Formula (\ref{equa8}). The red lines show that $T$ has a negative exponential relation with $\xi$, which is consistent with Formula (\ref{equa7}). That red lines marked with triangles are lower than red lines marked with squares show that IFP2 is faster than IFP1 on all six data sets.   
	\item[(2)] The blue lines show that $ERR$ scarcely changes when $\xi < 10^{-15}$. It seems both IFP1 and IFP2 have limitation in precision. We believe it is not a algorithm flaw, but caused by the insufficiency of C++'s DOUBLE data type, whose significant digit number is 15. $h_{i} < 10^{-15}$ can not change $\overline{\pi}_{i}$, thus $ERR$ stays constantly. 
\end{enumerate}

\subsection{Comparison with other algorithms}
\begin{figure}[htbp]
	\centering
	\includegraphics[width=0.32\textwidth,height=0.20\textheight]{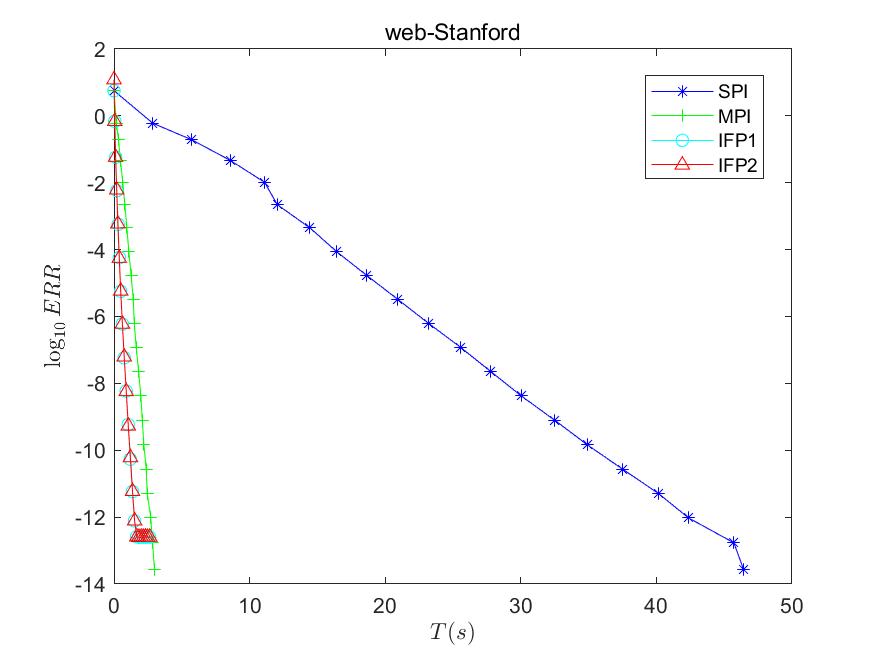}
    \includegraphics[width=0.32\textwidth,height=0.20\textheight]{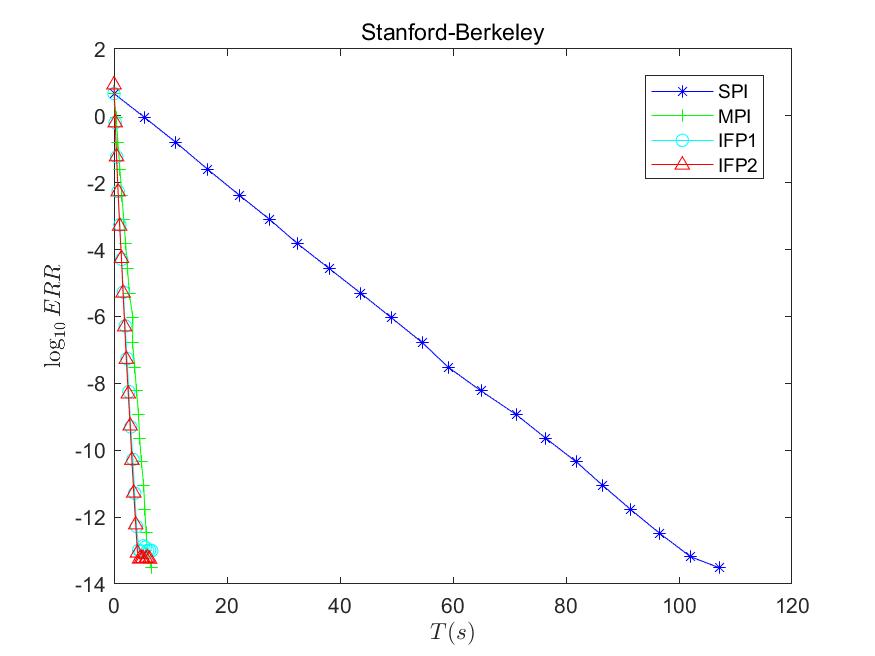}
    \includegraphics[width=0.32\textwidth,height=0.20\textheight]{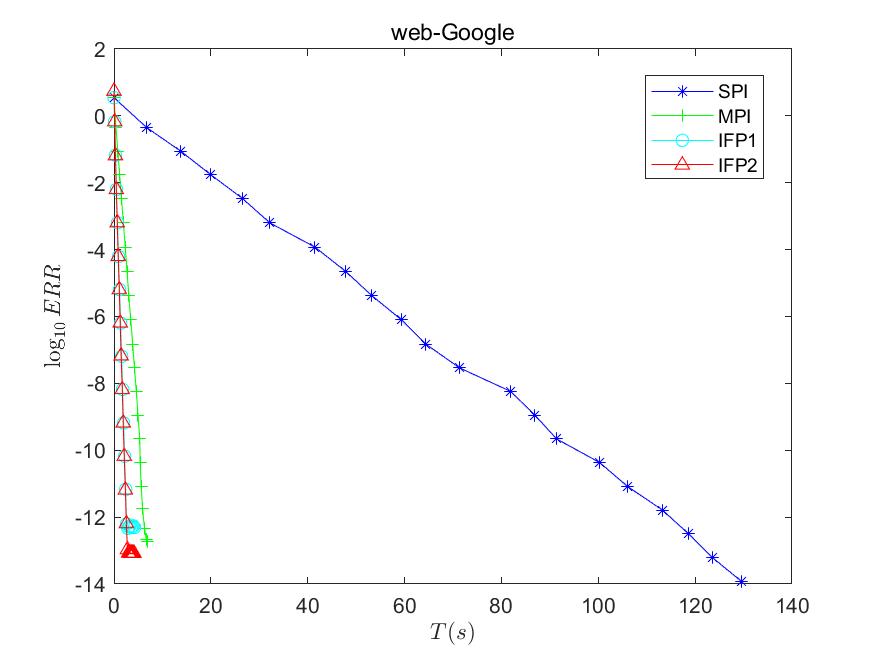}
    \includegraphics[width=0.32\textwidth,height=0.20\textheight]{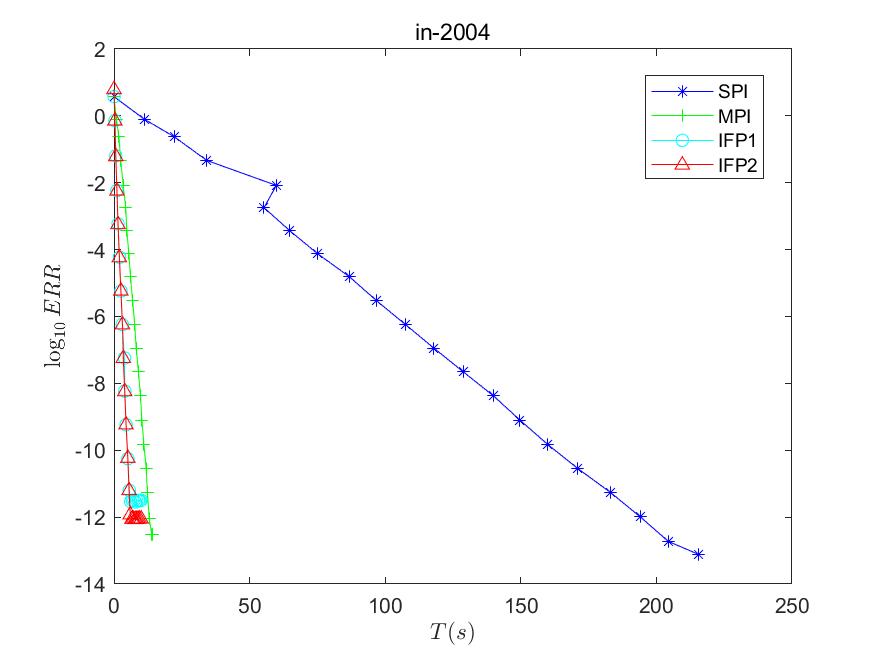}
    \includegraphics[width=0.32\textwidth,height=0.20\textheight]{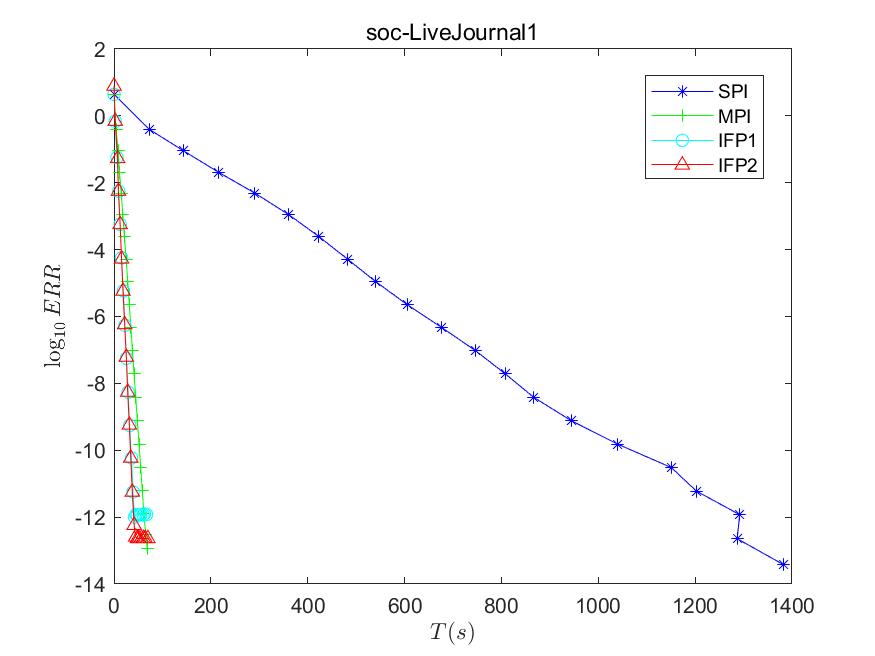}
    \includegraphics[width=0.32\textwidth,height=0.20\textheight]{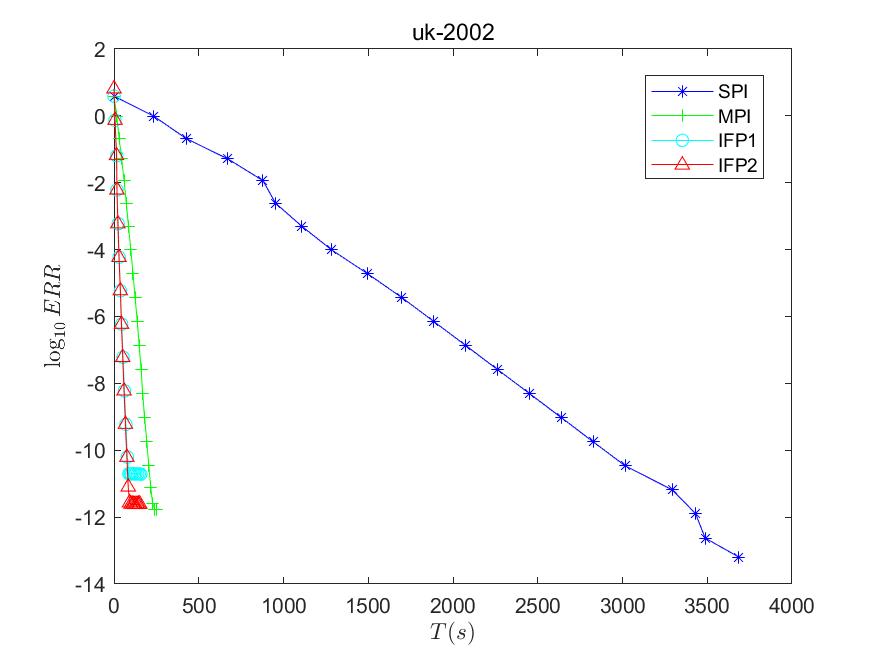}
	\caption{$T$ Versus $ERR$}
	\label{fig2}
\end{figure}

We compare IFP1 and IFP2 with Power method (SPI) \cite{brin2012reprint,page1999pagerank} and parallel Power method (MPI)\cite{duong2012parallel}. Both MPI, IFP1 and IFP2 are executed with 38 parallelism. The relation of $ERR$ and $T$ is illustrated in Figures \ref{fig2}. The CPU time consumption of preprocessing is illustrated in Table \ref{table2}. Table \ref{table3} illustrates the CPU time consumption $T$ when $ERR < 0.001$. 
\begin{enumerate}
	\item[(1)] The magenta, red and blue lines are lower than green lines in Figure \ref{fig2}, it shows that both MPI, IFP1 and IFP2 are faster than SPI, parallelizing is an effective solution to accelerate PageRank computing.
	\item[(2)] The magenta lines and red lines are lower than green lines in Figure \ref{fig2},  it shows that both IFP1 and IFP2 outperform MPI. Since IFP1, IFP2 and MPI are executed with the same parallelism, we believe the advantages owe to the higher convergence rate of IFP1 and IFP2. 
	\item[(3)] For MPI and IFP1, the preprocessing is assigning vertices to threads, for IFP2, the preprocessing includes assigning and invalidating the edges corresponding to the dangling vertices. MPI assigns all of the vertices to threads, thus consumes more time. The time consumption of invalidating is so few that it can be done in 0.155s even for uk-2002. 
	\item[(4)] Table \ref{table3} shows that IFP2 with 38 parallelism can be at most 50 times as fast as SPI, and at most 3 times faster than MPI, when $ERR<0.001$.  
\end{enumerate}

\begin{table}[htbp]
	\footnotesize
	\centering
	\begin{tabular}{l l l l}
		\toprule
		Data sets               &MPI       &IFP1        &IFP2    \\
		\midrule
		web-Stanford            &0.014     &0.012       &0.012   \\
		Stanford-Berkeley       &0.041     &0.030       &0.031    \\
		web-Google              &0.043     &0.033       &0.036   \\
		in-2004                 &0.070     &0.046       &0.054   \\
		soc-LiveJournal1        &0.236     &0.178       &0.194   \\
		uk-2002                 &0.926     &0.629       &0.784   \\
		\bottomrule
	\end{tabular}
	\caption{The time consumption of preprocessing}
	\label{table2}
\end{table}

\begin{table}[htbp]
	\footnotesize
	\centering
	\begin{tabular}{l l l l l}
		\toprule
		Data sets              &SPI            &MPI         &IFP1        &IFP2          \\
		\midrule
		web-Stanford           &14.445         &0.949       &0.303       &\textbf{0.285}    \\
		Stanford-Berkeley      &27.525         &1.709       &1.038       &\textbf{0.995}    \\
		web-Google             &32.182         &1.936       &0.718       &\textbf{0.671}    \\
		in-2004                &64.665         &4.728       &1.497       &\textbf{1.481}    \\
		soc-LiveJournal1       &423.183        &21.415      &12.702      &\textbf{12.617}   \\
		uk-2002                &1108.711       &85.455      &24.823      &\textbf{22.479}   \\
		\bottomrule
	\end{tabular}
	\caption{The time consumption when $ERR<0.001$}
	\label{table3}
\end{table}

\subsection{Performance under different parallelism}
\begin{figure}[htbp]
	\centering
	\includegraphics[width=0.32\textwidth,height=0.20\textheight]{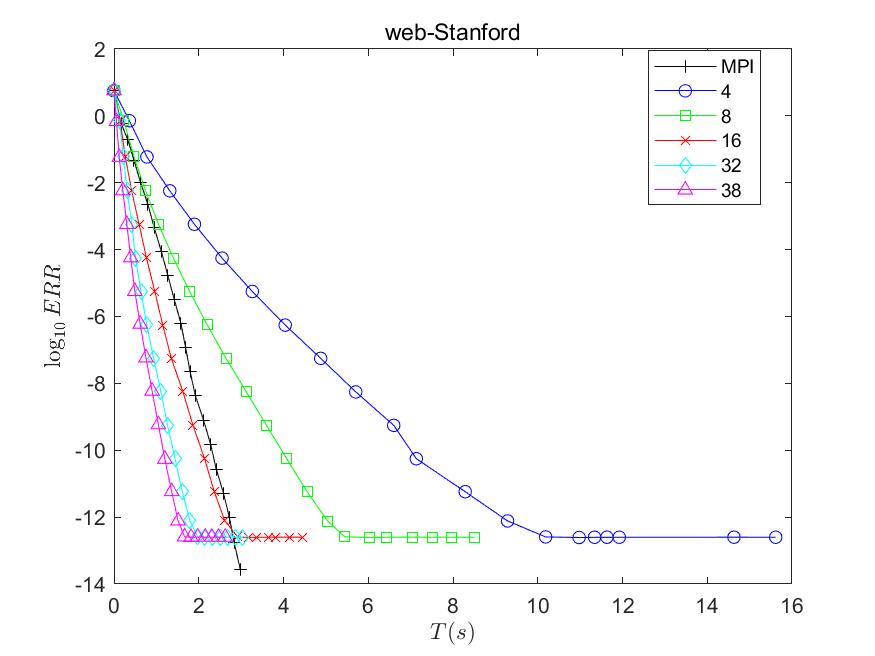}
	\includegraphics[width=0.32\textwidth,height=0.20\textheight]{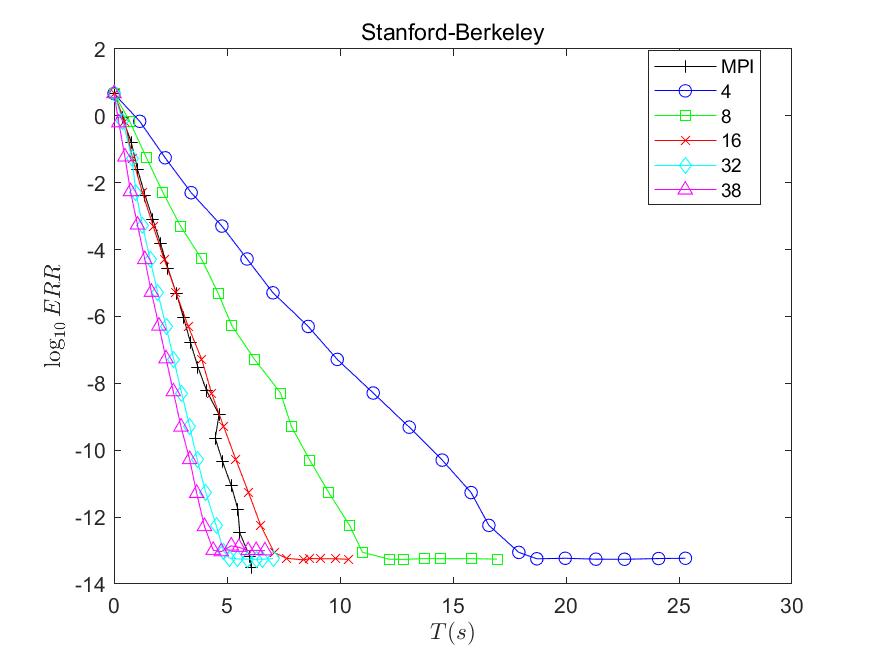}
	\includegraphics[width=0.32\textwidth,height=0.20\textheight]{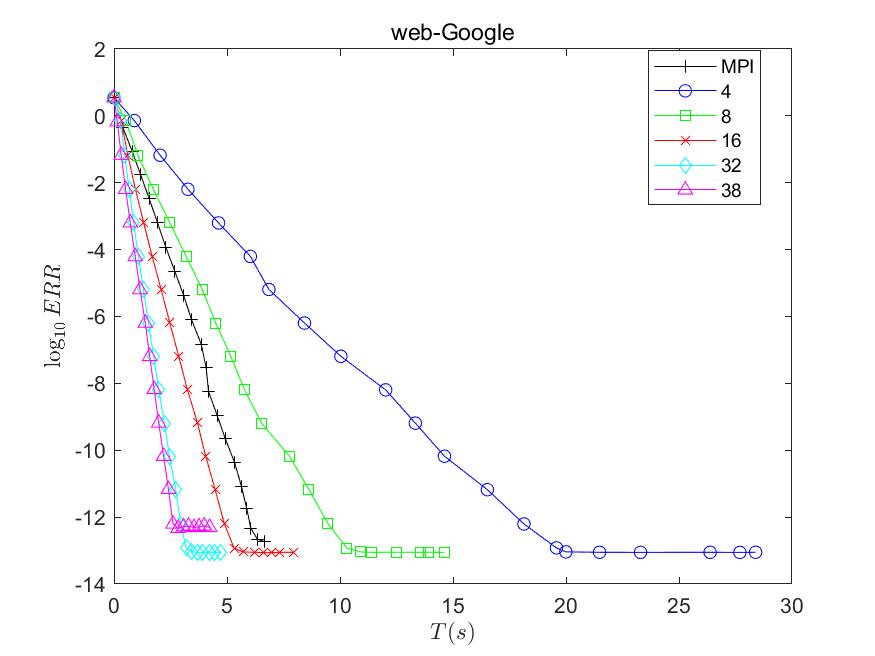}
	\includegraphics[width=0.32\textwidth,height=0.20\textheight]{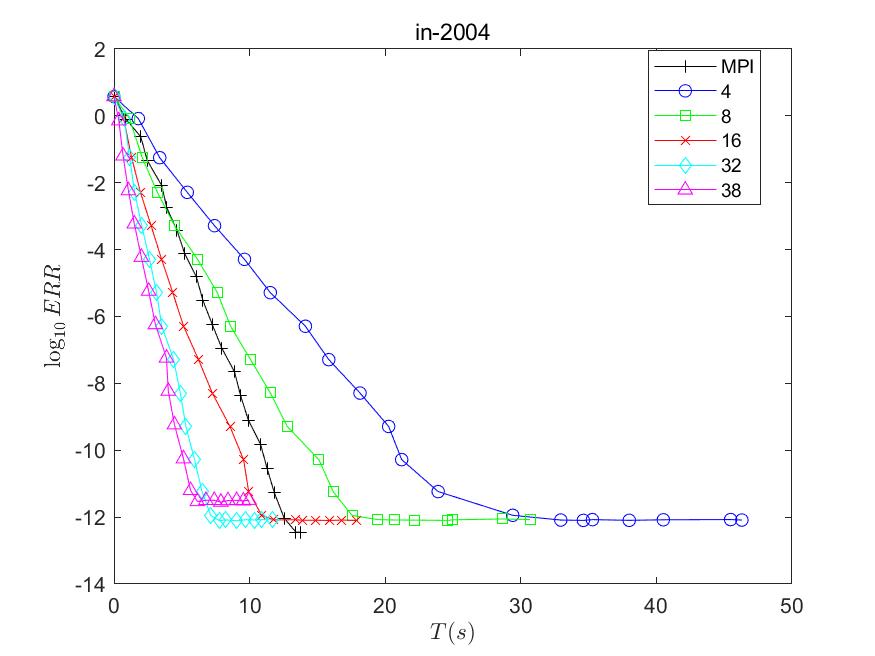}
	\includegraphics[width=0.32\textwidth,height=0.20\textheight]{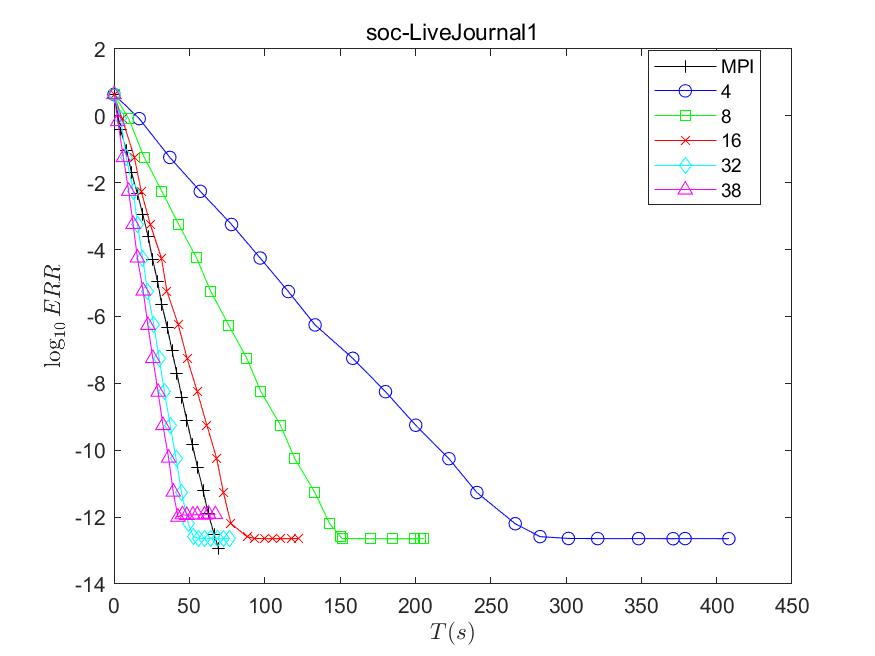}
	\includegraphics[width=0.32\textwidth,height=0.20\textheight]{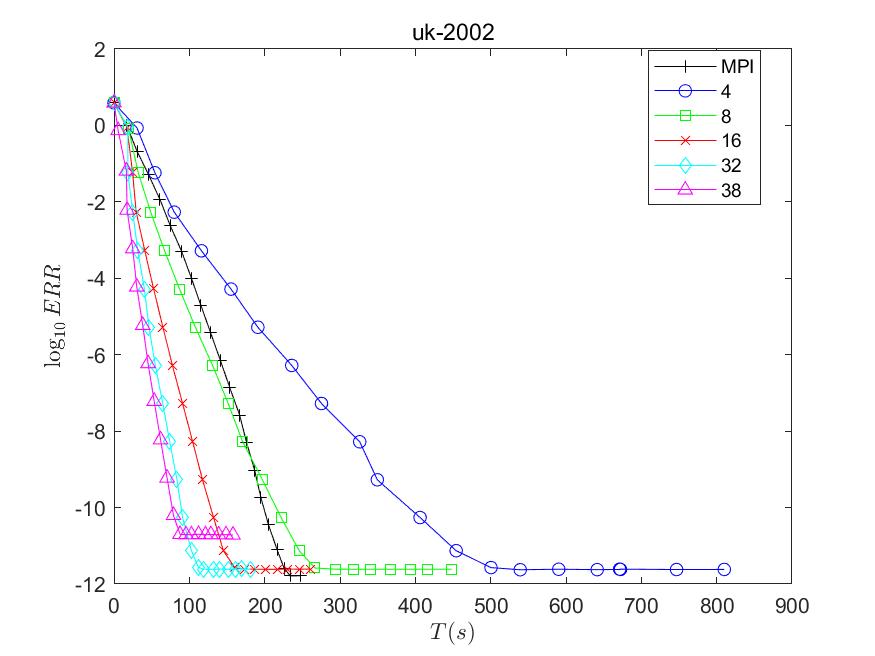}
	\caption{$T$ Versus $ERR$ under parallelism $4$,$8$,$16$,$32$ and $38$}
	\label{fig3}
\end{figure}

We execute IFP2 with parallelism 4, 8, 16, 32 and 38. The relation of $ERR$ and $T$ with different parallelism are illustrated in Figure \ref{fig3}. With the increasing of parallelism, the CPU time consumption $T$ decreases. Generally, the more the parallelism the faster the computing. IFP2 with 16 parallelism is faster than MPI with 38 parallelism, which proves IFP2 has higher convergence rate than Power method again. 

\section{Conclusion}\label{sec6}
PageRank is a basic problem of graph computation and parallelization is a feasible solution of accelerating computing. In this paper, we firstly reveal that PageRank vector is essentially the mass distribution, and base on which two parallel PageRank algorithms IFP1 and IFP2 are proposed. The most prominent feature of IFP1 is that it can make full use of the DAG structure, generally, the larger proportion of dangling vertices, the higher convergence rate, the larger proportion of unreferenced vertices and weak unreferenced vertices, the less computation amount. IFP2 pushes mass to the dangling vertices only once, thus, the more edges corresponding to the dangling vertices, the more computation decreases. Experiments on six data sets demonstrate that both IFP1 and IFP2 are faster than the Power method. However, every coin has two sides, on graph containing no DAG structure such as undirected graph, neither IFP1 nor IFP2 can outperform the Power method since the atomic operation is costly. In the future, PageRank on dynamic graph can be studied.

\bibliographystyle{elsarticle-num} 
\bibliography{ref}

\end{document}